\documentstyle[aps,preprint,epsfig]{revtex}

\begin{document}
\title{ Search for the scalar $a_0$ and $f_0$ mesons in the $\phi$ radiative
decays.}

\author{ N.N. Achasov and V.V. Gubin }
\address{ S.L. Sobolev Institute for Mathematics\\
630090 Novosibirsk 90, Russia}

\maketitle

\begin{abstract}
The potentialities of the production of the  $a_0$ and $f_0$ mesons in the 
$\phi$ radiative decays  are considered based on a new two-channel analysis
of the $\pi\pi$ scattering in a energy region near 1 GeV.  We predict 
 $BR(\phi\to\gamma f_0(980))\sim10^{-5}-10^{-4}$ that is a great value for 
the decay forbidden by the  Okubo-Zweig-Iizuka rule. We discuss the four-quark,
$K\bar K$-molecule and two-quark scenarios for the $a_0$ and $f_0$ mesons.
It is presented arguments that the $e^+e^-$-colliders and especially the $\phi$ 
factories give the possibility to choose a single one out of them.

\end{abstract}
\vspace{2cm}

The central problem of light hadron spectroscopy has been the problem
of the scalar $f_0(980)$ and $a_0(980)$ mesons. It is well known that these
states possess peculiar properties from the naive quark ($q\bar q$)
model point of view, see, for example \cite{achasov-84,achasov-91,achasov-1991}.
 To clarify the nature of these mesons a number
of models has been suggested.
It was shown that all their challenging properties
could be understood \cite{achasov-84,achasov-91,achasov-1991}
in the framework of the four-quark  ($q^2\bar q^2$) MIT-bag model
  with symbolic quark structure 
 $f_0(980)=s\bar s(u\bar u+d\bar d)/
\sqrt{2}$ and $a_0(980)=s\bar s(u\bar u-d\bar d)/\sqrt{2}$. Along with the
 $q^2\bar q^2$ nature of  $a_0(980)$ and $f_0(980)$ mesons the possibility of
 their being the  $K\bar K$ molecule is discussed.
During the last few years it was established
\cite{achasov-89,molecule,neutral} that the radiative decays
 of the $\phi$ meson $\phi\rightarrow\gamma f_0\rightarrow\gamma\pi\pi$ and
$\phi\rightarrow\gamma a_0\rightarrow\gamma\eta\pi$
could be a good guideline in distinguishing  the $f_0$ and $a_0$ meson
models. The branching ratios are considerably different in the cases of
naive quark, four-quark or molecular models. As has been shown
\cite{achasov-89,molecule,neutral}, in the four quark model the branching ratio is
\begin{equation}
BR(\phi\to\gamma f_0(q^2\bar q^2)\to\gamma\pi\pi)\simeq
BR(\phi\to\gamma a_0(q^2\bar q^2)\to\gamma\pi\eta)\sim10^{-4},
\end{equation}
and in the $K\bar K$ molecule model it is 
\begin{equation}
BR(\phi\to\gamma f_0(K\bar K)\to\gamma\pi\pi)\simeq
BR(\phi\to\gamma a_0(K\bar K)\to\gamma\pi\eta)\sim10^{-5}.
\end{equation}

 Currently also an interest in an old interpretation of the $f_0$
meson being an $s\bar s$ state is
rekindled, despite the fact that the almost ideal mass degeneracy of the
 $f_0$ and  $a_0$ mesons is difficult to understand in this case.
In spite of this fact, the  $s\bar s$ scenario is discussed in the current
literature as one possible model of the $f_0$ meson structure. 
It is easy to note that in the case of an $s\bar s$ structure
 of the $f_0$ meson
$BR(\phi\to\gamma f_0\to\gamma\pi\pi)$ and
$BR(\phi\to\gamma a_0\to\gamma\pi\eta)$ are different by factor of ten,
which should be visible experimentally.

 In the case when $f_0=s\bar s$ the suppression by the OZI rule is absent
and the evaluation gives \cite{achasov-89,neutral}
\begin{eqnarray}
BR(\phi\to\gamma f_0(s\bar s)\to\gamma\pi\pi)\simeq5\cdot10^{-5}, 
\end{eqnarray}
whereas for $a_0=(u\bar u-d\bar d)/\sqrt{2}$ the decay $\phi\to\gamma a_0\to
\gamma\pi\eta$ is suppressed by the OZI rule  and  is dominated by the real 
$K^+K^-$ intermidiate state breaking the OZI rule 
\cite{achasov-89,neutral}
\begin{eqnarray}
BR(\phi\to\gamma a_0(q\bar q)\to\gamma\pi\eta)\simeq(5\div8)\cdot10^{-6}.
\end{eqnarray}

Let us note that in the case of the $\phi\to\gamma\eta'$ decay allowed by the
OZI rule one expects  $BR(\phi\to\gamma \eta')\simeq(0.5\div1)\cdot10^{-4}$.

Imposing the appropriate photon energy cuts $\omega<100$ MeV, one can
show that the background reactions
$e^+e^-\to\rho(\omega)\to\pi^0\omega(\rho)\to\gamma\pi^0\pi^0$,
$e^+e^-\to\rho(\omega)\to\pi^0\omega(\rho)\to\gamma\pi^0\eta$ and
$e^+e^-\to\phi\to\pi^0\rho\to\gamma\pi^0\pi^0(\eta)$ are negligible 
in comparison with the scalar meson contribution
 $e^+e^-\to\phi\to\gamma f_0(a_0)\to\gamma\pi^0\pi^0(\eta)$
for  $BR(\phi\to\gamma f_0(a_0)\to\gamma\pi^0\pi^0(\eta))$
greater than $5\cdot10^{-6}(10^{-5})$.

Let us consider the reaction $e^+e^-\to\phi\to\gamma (f_0+\sigma)\to\gamma
\pi^0\pi^0$ with regard to the mixing of the $f_0$ and $\sigma$ mesons.
We consider the one loop mechanism of the $R$ meson production, where
$R=f_0,\sigma$, through the charged kaon loop, $\phi\to K^+K^-\to\gamma R$,
see \cite{{achasov-89},{molecule},{neutral}}. 
The whole formalism in the frame of which we study this problem is 
discussed in \cite{neutral}. The parameters of the $f_0$ and $\sigma$
mesons we obtain from fitting the $\pi\pi$ scattering data, see 
\cite{neutral}.

In the four-quark model  we consider the following parameters  to be   free:
the coupling constant of the $f_0$ meson to the $K\bar K$
channel $g_{f_0K^+K^-}$, the coupling constant of the $\sigma$ meson
to the $\pi\pi$ channel $g_{\sigma\pi\pi}$, the constant of the
$f_0-\sigma$ transition  $C_{f_0\sigma}$,
the ratio $R=g^2_{f_0K^+K^-}/g^2_{f_0\pi^+\pi^-}$, the phase $\theta$ of the
elastic background and the $\sigma$ meson mass. The mass of the $f_0$ meson
is restricted to the region $0.97<m_{f_0}<0.99$ GeV.  We treat the  $\sigma$
meson as an ordinary two-quark state 
 $g_{\sigma K^+K^-}=g_{\sigma\pi^+\pi^-}/2$.
One gets $g_{\sigma K^+K^-}=\sqrt{\lambda}
 g_{\sigma\pi^+\pi^-}/2\simeq0.35g_{\sigma\pi^+\pi^-}$.  So the constant
$g_{\sigma K^+K^-}$ ( and   $g_{\sigma\eta\eta}$ )  is not essential in 
our fit. 

As for the reaction $e^+e^-\to\gamma\pi^0\eta$  the similar analyisis
of the $\pi\eta$ scattering cannot be performed directly. But, our analysis
of the final state  interaction for the $f_0$ meson production show that
the situation does not changed radically, in any case in the region
$\omega<100$ MeV. Hence, one can hope that the final state interaction
in the  $e^+e^-\to\gamma a_0\to\gamma\pi\eta$ reaction will not strongly
affect the predictions in the region $\omega<100$ MeV. 
Based on the analysis of $\pi\pi$ we predict the quantities of the
 $BR(\phi\to\gamma a_0\to\gamma\pi\eta)$ in the  $q^2\bar q^2$ model, 
 $K\bar K$ model and the $q\bar q$ model where  $f_0=s\bar s$ and
$a_0=(u\bar u-d\bar d)/\sqrt{2}$.

The fitting shows that in the four quark model
($g^2_{f_0K^+K^-}/4\pi >1\ GeV^2$ ) a number of parameters 
describe well enough the $\pi\pi$ scattering in the region $0.7<m<1.8$ GeV. 
We predict 
$BR(\phi\to\gamma (f_0+\sigma)\to\gamma\pi\pi)\sim10^{-4}$ and
$BR(\phi\to\gamma a_0\to\gamma\pi\eta)\sim10^{-4}$ in the $q^2\bar q^2$
model.

In the model of the $K\bar K$ molecule we get
$BR(\phi\to\gamma (f_0+\sigma)\to\gamma\pi\pi)\sim10^{-5}$ and
$BR(\phi\to\gamma a_0\to\gamma\pi\eta)\sim10^{-5}$.

In the $q\bar q$ model the $f_0(a_0)$ meson is considered as a
point-like object, i.e. in the  $K\bar K$ loop, $\phi\to K^+K^-\to\gamma
f_0(a_0)$ and in the transitions caused by the $f_0-\sigma$ mixing we consider
both the real and the virtual intermediate states. This model is different from
$q^2\bar q^2$ model by the coupling constant which is  $g^2_{f_0K^+K^-}/4\pi
<0.5\ GeV^2$.  In this model  we obtain
$BR(\phi\to\gamma (f_0+\sigma)\to\gamma\pi\pi)\simeq5\cdot10^{-5}$ and
taking into account the imaginary part of the decay amplitude only,
as the main one, we get
$BR(\phi\to\gamma a_0(q\bar q)\to\gamma\pi\eta)\simeq8\cdot10^{-6}$.

Notice that we could not find the parameters at which
the data on the $\pi\pi$ scattering are described well but the
$BR(\phi\to\gamma (f_0+\sigma)\to\gamma\pi\pi)$ is less then $10^{-5}$.
Or, more precisely, we could get  $BR(\phi\to\gamma (f_0+\sigma)
\to\gamma\pi\pi)\simeq10^{-5}$, only if we droped out the real part of the
$K^+K^-$ loop in the $\phi\to K^+K^-\to\gamma (f_0+\sigma)$ decay amplitude,
 otherwise  we got
$BR(\phi\to\gamma (f_0+\sigma)\to\gamma\pi\pi)>4.5\cdot10^{-5}$.
 Forgetting the possible models, one can say that from
the $\pi\pi$ scattering results
 $BR(\phi\to\gamma (f_0+\sigma)\to\gamma\pi\pi)>10^{-5}$,
  that is enormous for the suppressed by the OZI rule decay. 

The experimental data from SND detector, submited to this conference,
support the four quark nature
of the $f_0$  and $ a_0$  mesons, see \cite{snd1,snd2}.

The figures bellow show the results of fitting in the $q^2\bar q^2$ model 
for the parameters:
$\theta=60^{\circ}$, $R=2.0$, $g^2_{f_0K^+K^-}/4\pi=0.72\ GeV^2$,
$g^2_{\sigma\pi\pi}/4\pi=1.76\ GeV^2$, $C_{f_0\sigma}=-0.17\ GeV^2$,
$m_{\sigma}=1.47\ GeV$, $m_{f_0}=0.98\ GeV$. The effective width of the
 $f_0$ meson $\Gamma_{eff}\simeq60\ MeV$. The $BR_{f_0+\sigma}(BR_{f_0})=
3\cdot BR(\phi\to\gamma(f_0+\sigma)\to\gamma\pi^0\pi^0)(
3\cdot BR(\phi\to\gamma f_0\to\gamma\pi^0\pi^0))=1.18(1.43)\cdot10^{-4}$
at $\omega<250\ MeV$ and  $BR_{f_0+\sigma}(BR_{f_0})=0.63(0.6)\cdot10^{-4}$
at $\omega<100\ MeV$.

\begin{figure}[b!] 
\centerline{\epsfig{file=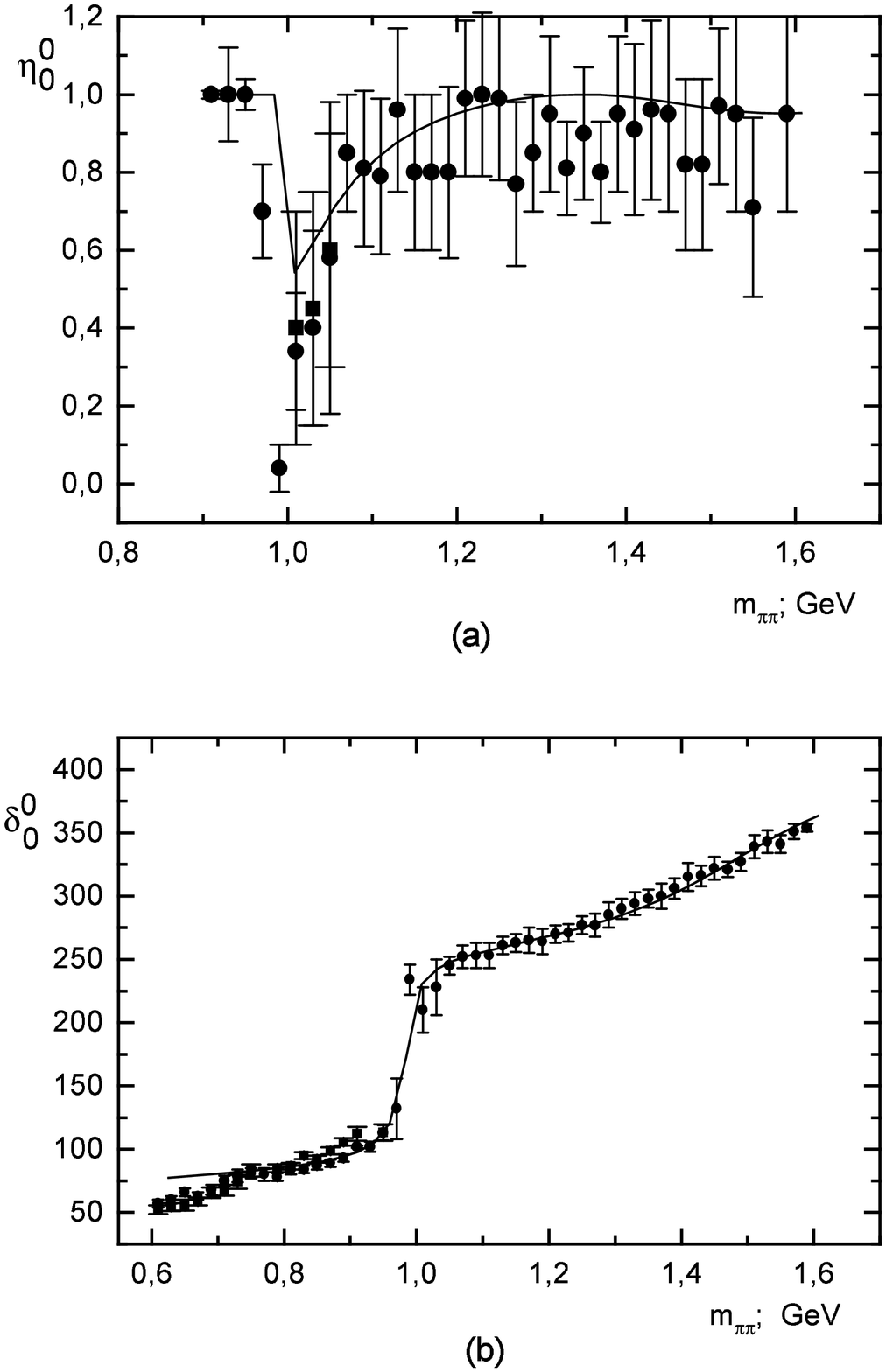,height=3.5in,width=3.5in}}
\vspace{10pt}
\caption{(a) The inelasticity $\eta^{I=0}_{L=0}$. (b) 
The phase $\delta^{I=0}_{L=0}$. }

\end{figure}

\begin{figure}[b!] 
\centerline{\epsfig{file=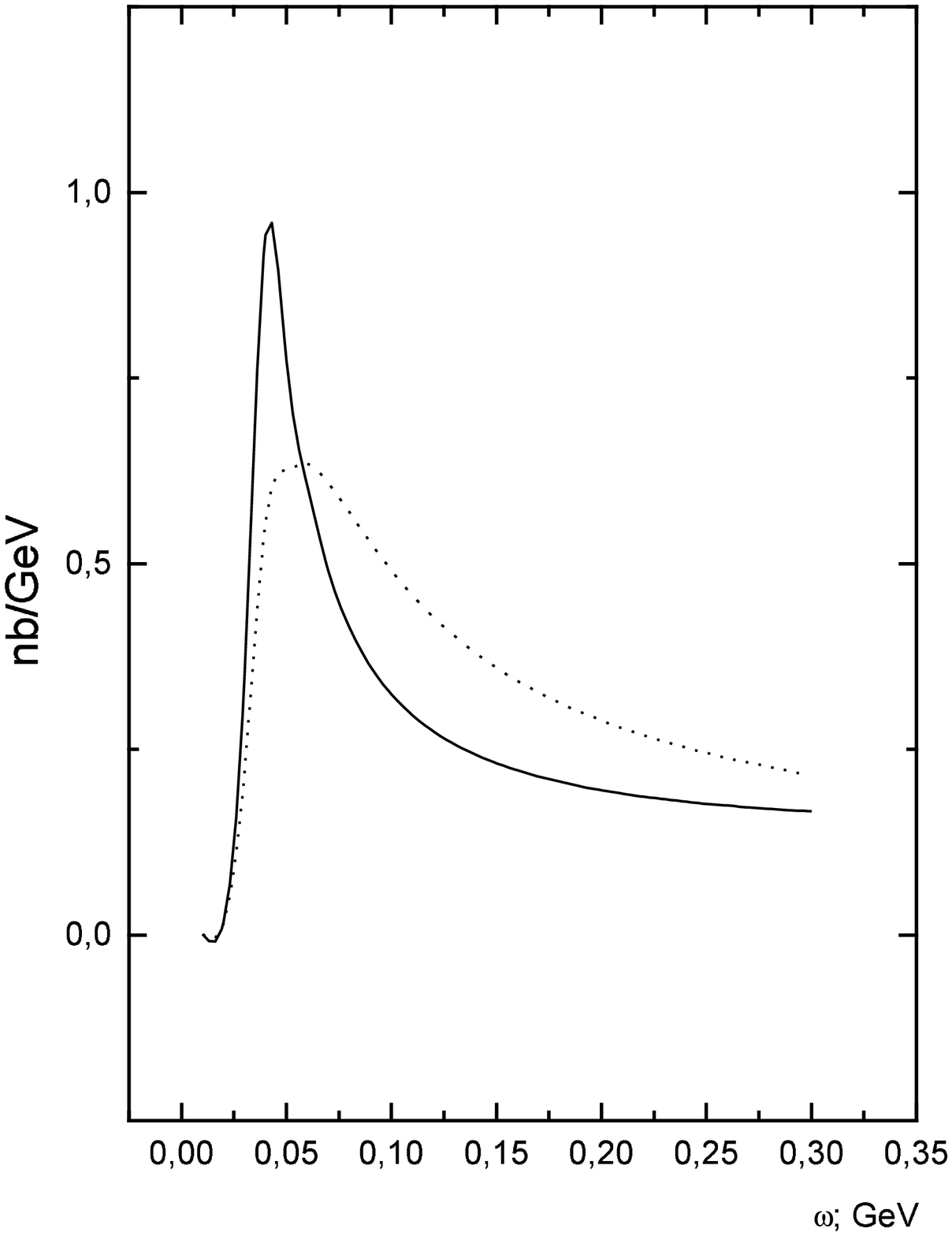,height=2.5in,width=2.5in}}
\vspace{10pt}
\caption{ (c) The spectrum of the differential cross
section $d\sigma(e^+e^-\to\gamma(f_0+\sigma)\to\gamma\pi^0\pi^0)/d\omega$ 
with  mixing of the $f_0$ ang $\sigma$ mesons).  
The dashed line is the spectrum of the $f_0$ meson without mixing with
the $\sigma$ meson. }
\end{figure}

\end{document}